\documentclass[12pt]{article} \input epsf.tex
\usepackage{epsfig,graphicx,}
\usepackage{amsmath,amsfonts,amssymb,stmaryrd,latexsym}
\usepackage{hyperref,cite}
\usepackage{subcaption,comment} 
\newcommand{\be}{\begin{equation}}
\newcommand{\ee}{\end{equation}}
\newcommand{\bea}{\begin{eqnarray}}
\newcommand{\eea}{\end{eqnarray}}
\newcommand{\bear}{\begin{eqnarray}}
\newcommand{\eear}{\end{eqnarray}}
\newcommand{\ba}{\begin{array}}
\newcommand{\ea}{\end{array}}

\newcommand{\eg}{{\it e.g.}}
\newcommand{\ie}{{\it i.e.}}

\newcommand{\gammah}{h^0 \gamma}

\usepackage[usenames,dvipsnames]{xcolor}

\usepackage{dcolumn}
\usepackage{slashed}

\newcommand{\lsim}{\!\mathrel{\hbox{\rlap{\lower.55ex \hbox{$\sim$}} \kern-.34em \raise.4ex \hbox{$<$}}}}
\newcommand{\gsim}{\!\mathrel{\hbox{\rlap{\lower.55ex \hbox{$\sim$}} \kern-.34em \raise.4ex \hbox{$>$}}}}

\def\GeV{\text{ GeV}}

\def\fb{\text{ fb}}

\newcommand{\Fref}[1]{Figure~\ref{#1}}
\newcolumntype{M}{>{$}c<{$}}

\begin{document}

\baselineskip=18pt \pagestyle{plain} \setcounter{page}{1}

\vspace*{-2.1cm}

\noindent \makebox[11.5cm][l]{\small \hspace*{-.2cm} }{\small Fermilab-Pub-17-169-T}  \\  [-1mm]

\begin{center}

{\large \bf   Higgs-photon resonances } \\ [9mm]

{\normalsize \bf Bogdan A. Dobrescu, Patrick J. Fox and John Kearney \\ [3mm]
{\small {\it
Theoretical Physics Department, Fermilab, Batavia, IL 60510, USA   
  }}\\
}

\center{May 23, 2017}

\end{center}

\vspace*{0.2cm}

\begin{abstract}
We study models that produce a Higgs boson plus photon ($h^0 \gamma$) resonance at the LHC.
When the resonance is a $Z'$ boson, decays to $h^0 \gamma$ occur at one loop. If the $Z'$ boson couples
at tree-level to quarks, then the $h^0 \gamma$ branching fraction is typically of order  $10^{-5}$ or 
smaller. Nevertheless, there are models that would allow the observation of $Z' \to h^0 \gamma$
at $\sqrt{s} = 13$ TeV with a  cross section times branching fraction larger than 1 fb for a $Z'$ mass 
in the 200--450 GeV range, and larger than 0.1 fb for a mass up to 800 GeV. 
The 1-loop decay of the $Z'$ into lepton pairs competes with $h^0 \gamma$, even if the $Z'$ couplings to 
leptons vanish at tree level.
We also present a model in which a $Z'$ boson decays into a Higgs boson and a pair of collimated photons, mimicking 
an $h^0 \gamma$ resonance.  In this model, the $h^0 \gamma$ resonance search would be the discovery mode for 
a $Z'$ as heavy as 2 TeV. 
When the resonance is a scalar, although decay to $h^0 \gamma$ is forbidden by angular momentum conservation,
the  $h^0$ plus collimated photons channel  is allowed.
We comment on prospects of observing an $h^0 \gamma$  resonance through different 
Higgs decays, on constraints from related searches, and on models where $h^0$ is replaced by a nonstandard Higgs boson.
\end{abstract}

\vspace*{0.5cm}

\newpage 

{\small
\tableofcontents
}

\section{Introduction} \setcounter{equation}{0}

The ATLAS and CMS experiments at the Large Hadron Collider (LHC) are searching for new particles and interactions in a large number of final states.
Among these, a particularly clean class of 
probes is the resonant production of two Standard Model (SM) particles. A signal of this type would indicate the existence of a
new particle that has 2-body decays. Searches for 2-body resonances have covered many combinations of SM particles. 
Nevertheless, there are some combinations of two SM particles that remain to be searched for at the LHC \cite{Craig:2016rqv}.
Existing resonant searches that involve the Higgs boson, $h^0$, and another SM particle in the final state include only $h^0W$ and $h^0 Z$ \cite{Aad:2015yza, Khachatryan:2016yji, Khachatryan:2015bma}. 

In this paper we study theoretical and phenomenological constraints on resonances that consist of a Higgs boson and a photon.
A particle that can decay into $h^0 \gamma$ must be a boson. Furthermore, angular momentum conservation prevents that particle from having spin 0.  
A simple way to prove that a spin-0 particle cannot decay into another spin-0 particle and a photon is to show that the decay amplitude vanishes for 
any operators that involve these three fields.

Thus, the leading candidate for a $h^0 \gamma$ resonance is a spin-1 particle, usually known as a $Z'$ boson (as it has to be electrically neutral and color singlet).
Electromagnetic gauge invariance allows the  $Z' \to h^0 \gamma$ process only through higher-dimensional operators, which arise from loops. 
Therefore, this partial width is many orders of magnitude smaller than the $Z'$ mass. If the $Z'$ has sizable couplings to quarks, as required to ensure large  $Z'$ production 
at the LHC, then the $B(Z' \to h^0 \gamma)$ branching fraction is very small. We give examples of renormalizable  $Z'$ models, and we compute this $h^0 \gamma$ branching fraction.
We will find that $B(Z' \to h^0 \gamma)$ is typically of the order of $10^{-5}$ or smaller. Nevertheless,
the observation of the $Z'$ in this mode is possible for a range of parameters, due to the small backgrounds. 

The signal for a ``Higgs-photon" resonance can be much larger than in the case of the loop-induced process if what appears to be the photon is in fact a cluster of photons. 
Consider a heavy boson that decays into a Higgs boson and a spin-0 particle, $A^0$, of GeV-scale mass. If $A^0$ subsequently decays into a pair of photons, then 
the large boost of $A^0$ in the lab frame makes the two photons overlap in the electromagnetic calorimeter,  leading to a single-photon signature \cite{Dobrescu:2000jt}. 
The heavy boson in this case may be a $Z'$, or even a spin-0 particle given that its decay into $h^0 A^0$ conserves angular momentum.
Eventually, with more detailed studies and larger data sets, the collimated photons (collectively labelled by $``\gamma"$) may be distinguished from a single photon.

In Section 2 we discuss phenomenological issues associated with possible Higgs-photon resonance searches, and estimate the current cross-section limits 
based on related experimental results. The $Z'$ models and their predictions for resonant searches at the LHC are presented in Section 3. The
collimated photon scenario is discussed in Section 4. There, we also describe a renormalizable model that leads to 
$Z' \to h^0 A^0 \to h^0+\! ``\gamma"$.
Section 5 includes our conclusions as well as some comments on $h' \gamma$ resonances, where $h'$ is a new Higgs-like boson.

\bigskip

\section{Prospects for Higgs-photon resonance searches}
\label{sec:exptprospects}

There are currently no published searches for $\gammah$ resonances. Yet, such searches could provide an interesting test of physics beyond the SM.
In this section we discuss the prospects for such a search at the LHC, with particular focus on the current experimental reach that could be achieved with minimal extension of pre-existing 
searches in related channels.

We concentrate on the final state in which the Higgs boson decays into bottom quarks.
We will denote the decaying particle by  $Z'$ in this section, as for a new vector boson, but our results are more generally applicable (the case of a spin-0 resonance is briefly discussed in Section 4).
First, consider a light resonance, $M_{Z'} \lsim 700 \GeV$, so that the final state is $b\bar{b}\gamma$ with resolved $b$ jets. The dominant SM background is non-resonant $b\bar b \gamma$ 
production and, to a lesser extent, $jj\gamma$ with mis-tags. Requiring that $110\GeV < m_{b\bar{b}} < 135 \GeV$ and that $p_T(\gamma)>50 \GeV$ gives a background cross section at 
the 13 TeV LHC of $\sim 0.5$ pb (based on simulation with MadGraph \cite{Alwall:2014hca}), peaked at low $m_{b\bar{b}\gamma}$ invariant mass.
The large background will make a search in this regime challenging. However, various kinematic features of the signal, including the resonant peak in $m_{b\bar{b}\gamma}$, can be used to differentiate the signal
from the background. With 3000 fb$^{-1}$ of data, a signal cross sections even below $\sim 1$ fb can be observed.
Moreover, as we discuss in the context of specific models below, low-mass $\gammah$ resonances 
may offer the best prospects for observation due to the larger rates. This makes the development of a dedicated search strategy for such resonances particularly important.

For a heavier $Z'$ boson, $M_{Z'} \gsim 700 \GeV$, the Higgs boson will be boosted and the two $b$ quarks will be contained within a jet of size $\Delta R \sim 2M_h/p_{T, h} \lsim 0.7$.  Thus, the observed final state is a photon and a wide jet with substructure.
As the final state ($\gamma j$) is similar to that considered in searches for excited quarks, we can reinterpret the results of these searches to estimate 
the current and future reach of a dedicated $\gammah$ search in this regime.

The dominant backgrounds for the excited quark search are continuum $\gamma j$ as well as QCD with jet misidentification \cite{Khachatryan:2014aka,CMS:2016qtb,Aad:2013cva}.  However, these backgrounds can be efficiently suppressed in the case of $\gammah$ by applying a Higgs tagger to the jet, as for the $Wh^0$ resonance search \cite{Khachatryan:2016yji}.  
The Higgs tagger requires the jet mass lie close to the Higgs mass, as well as the presence of a heavy flavor tag in the jet, and is $\simeq 80\%$ efficient while having a $\simeq 10\%$ jet misidentification rate \cite{Khachatryan:2016yji}.  Estimating the effect of applying the Higgs tagger directly to the existing excited quark search is complicated by the fact that the tagger is based on Cambridge-Aachen jets with $\Delta R = 0.8$ while jets in the excited quark search are clustered according to the anti-$k_T$ algorithm with $\Delta R = 0.5$.  Increasing the jet radius leads to more background jets passing the cut $p_{T, j} > 170 \GeV$ applied in \cite{Khachatryan:2014aka}. 
We estimate this increase to be around 10\%, by simulating the dominant direct background $p p \rightarrow \gamma j$ in MadGraph \cite{Alwall:2014hca}, with showering carried out subsequently in \textsc{Pythia 6.4} \cite{Sjostrand:2006za} and detector simulation using \textsc{Delphes 3.3.0} \cite{deFavereau:2013fsa}. Jets are clustered both ways using FastJet \cite{Cacciari:2011ma}, and the proportion of events passing the experimental cuts is used to estimate the impact of changing jet algorithms.
Overall, the effect of increasing the jet radius and applying the Higgs tagger is expected to reduce the background by a factor of approximately 10. .

The CMS excited quark search \cite{Khachatryan:2014aka} does not present the acceptance times efficiency for the $q^*\rightarrow q\gamma$ signal to pass the analysis, so we must estimate this quantity.  For a given $q^*$ resonance mass $M$ we estimate the number of expected background events with $|m_{\gamma j}-M|<0.12 M$, which corresponds to a $3\sigma$ window for the CMS resolution, using the functional form for the background provided in \cite{Khachatryan:2014aka}.  By comparing our estimated limits, based on these background estimates, with those published by the collaboration, we determine an approximately mass-independent acceptance times efficiency, ${\cal A}\,  \epsilon \approx 0.5$, for $q^*$ signal events.

\begin{figure}[t] 
\vspace*{2mm}
   \centering
    \includegraphics[width=0.65\textwidth, angle=0]{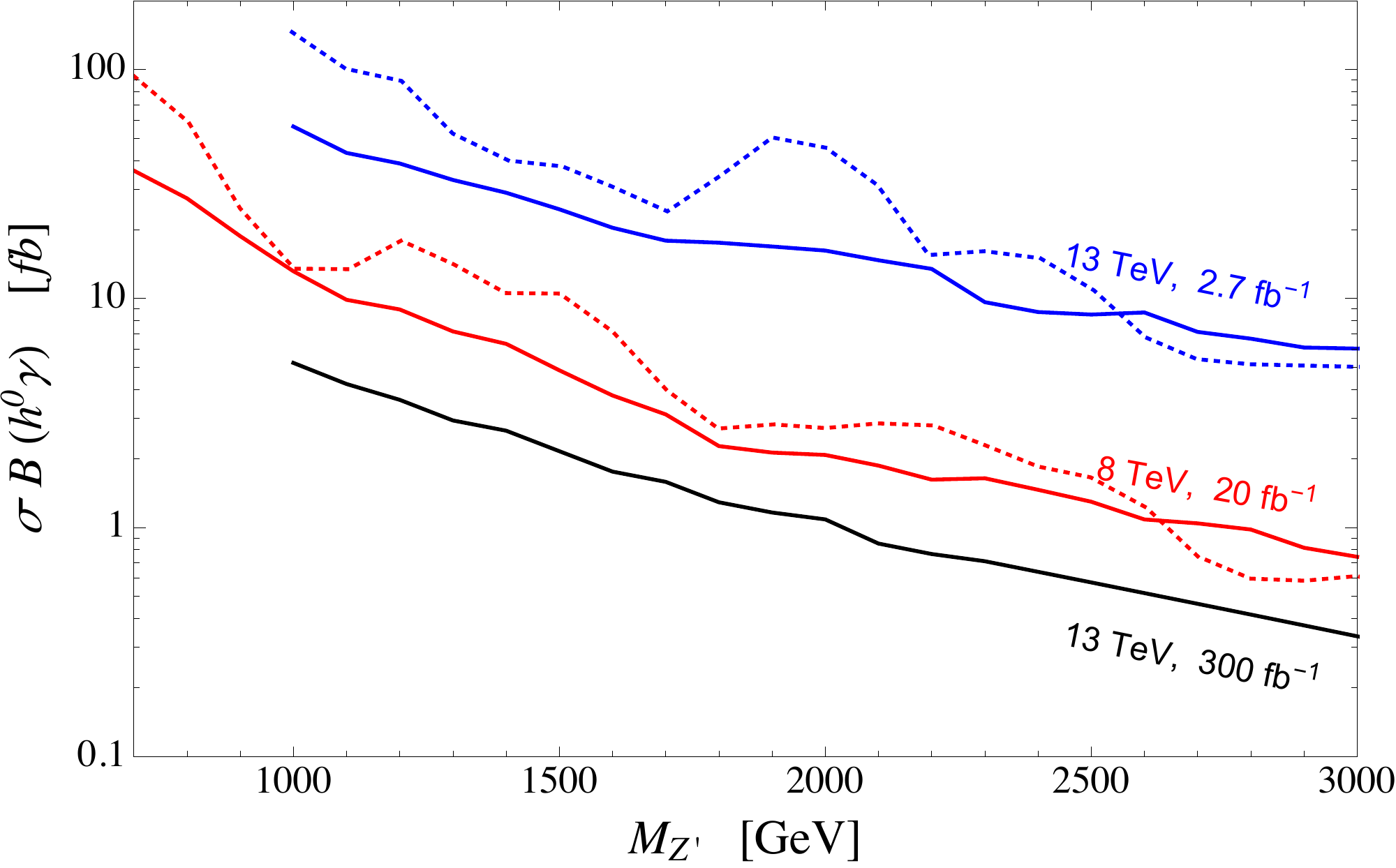}
   \caption{Estimated limits on the cross section for an 
   $\gammah$  resonance obtained by recasting the $pp \to q^\ast \rightarrow \gamma q$ search results:
current limits at $\sqrt{s} = 13$ TeV and 8 TeV (solid blue and red lines, respectively), and projected limit at 13 TeV with ${\cal L} = 300 \fb^{-1}$ (solid black line). 
For comparison, dotted lines represent the current CMS \cite{Khachatryan:2014aka,CMS:2016qtb} limits  on the cross section times branching fraction for an excited quark decaying $q^\ast \rightarrow \gamma q$; 
the improvement by a factor of approximately 1.5 over much of the mass range is the result of applying the Higgs tagger.}
   \label{fig:gammahlimits}
\end{figure}

Assuming that ${\cal A} \, \epsilon$ for our signal is the same as that derived for an excited quark we estimate the reach of an excited quark search applied to an $\gammah$ resonance.  Due to changes in jet size and mistagging rates the background is rescaled by $\simeq 0.11$, and the signal rate is given by $0.8 {\cal A} \, \epsilon \,  \sigma B (Z' \rightarrow \gammah) B (h^0 \rightarrow b \bar{b})$.  Using excited quark searches both at 8 TeV with ${\cal L} = 20 \fb^{-1}$ \cite{Khachatryan:2014aka} and 13 TeV with ${\cal L} = 2.7 \fb^{-1}$ \cite{CMS:2016qtb}, we project the current limits on an $\gammah$ resonance that would be obtained were a search similar to that described here implemented by the CMS collaboration. We also use these limits to project the reach that could be obtained at 13 TeV with ${\cal L} = 300 \fb^{-1}$. These limits are shown in \Fref{fig:gammahlimits}.
Thus, we estimate that the cross-section sensitivity to an  $\gammah$ resonance in the mass range 1--3 TeV at the 
8 TeV LHC ranges from 10 fb to 1 fb, and at the 13 TeV LHC with 300 fb$^{-1}$ of data will be ranging from 5 fb to 0.5 fb.
Comparing the 8 TeV and 13 TeV reach, the background increases by a factor of about $3$, while  the increase in the signal cross section depends on the initial state, which is model dependent.

Additional Higgs decays can be used to search for $\gammah$ resonances. 
For example, a search in the a photon-plus-lepton final state would be sensitive to $h^0 \to WW^* \to \ell\nu jj$  and  $h^0 \to  \tau^+\tau^-$. 
Also, the production of an $\gammah$ resonance followed by the Higgs boson decays into photon pairs would lead to a spectacular 
$3\gamma$ peak, that would allow a precise determination of the resonance mass.  There is an ATLAS search for a $3\gamma$ resonance \cite{Aad:2015bua}, 
but it is not sensitive enough to be applied to $\gammah$ resonances because the Higgs branching fraction into photons is too small. 

Other $Z'$ decay channels may be important.
The relative branching fractions and the search sensitivities of various channels will determine whether $\gammah$ represents a potential discovery channel for the new resonance or a precision probe of a resonance discovered in another channel. The possible $Z'$ decays are model dependent, but certain other channels may be particularly relevant.
For instance, as the $Z'$ must be produced in hadron collisions, it is likely to have a sizeable branching fraction to jets. Meanwhile, for a vector resonance, the $Z'$ may also be able to decay to lepton pairs, as will be discussed in the next section.
Comparing the estimated 8 TeV sensitivity to $\gammah$ with those to dijet \cite{Khachatryan:2015sja} and dilepton \cite{Aad:2014cka} resonances, we find that the dijet channel is 10--100 times less sensitive in the high mass regime, while the dilepton channel is approximately 10--30 times more sensitive. 
Below, we will discuss the impact of these other channels, and their importance for interpreting the results of an $\gammah$ search in specific models.

%
\section{Vector boson decays to $\gammah$} \setcounter{equation}{0}
\label{sec:vectorhgamma}
\setcounter{equation}{0}

A new vector boson $Z'$ can decay to $\gammah$ via an  operator of the form
\be
\label{eq:vectordecayop}
\frac{c_\gamma \, e \,  v}{(4\pi)^2 \, m_0^2} \, h Z'_{\mu \nu} F^{\mu \nu}  ~~,
\ee
where $Z'_{\mu \nu} = \partial_\mu Z'_\nu -  \partial_\nu Z'_\mu $, and $F_{\mu \nu}$ is the electromagnetic field strength;
$m_0$ is the mass of some particle running in the loop,
$e \approx 0.3$ is the electromagnetic gauge coupling, $v \approx 246$ GeV is the weak scale. 
The  factor of $(4\pi)^2$ in the denominator is associated with the loop integral, so that
the model-dependent dimensionless parameter $c_\gamma$ is typically of order one or smaller. 

An additional operator, $h \widetilde Z'_{\mu \nu} F^{\mu \nu}$, can also contribute to $Z' \to \gammah$. However, the coefficient of
that operator vanishes in the limit of CP conservation, and we will ignore it here.  
$U(1)_{\rm em}$ gauge invariance ensures any additional operators (including higher dimension operators) 
contributing to on-shell decay $Z' \rightarrow \gammah$ can be related to operator (\ref{eq:vectordecayop}) using the equations of motion.
A straightforward way to see this is to consider the matrix element for the decay. The Ward identity requires $p_\mu (\gamma) \, \epsilon_\nu (Z') {\cal M}^{\mu \nu} = 0$, where the full matrix element is ${\cal M} = \epsilon_\mu^*(\gamma) \, \epsilon_\nu(Z') {\cal M}^{\mu\nu}$ with $\epsilon$ the polarization of the 
$Z'$ boson or photon. So, writing ${\cal M}^{\mu \nu}$ in terms of momenta, it must be the case that
\be
{\cal M}^{\mu \nu} \propto p^\mu\!(Z')\, p^\nu \!(\gamma) - p(Z') \cdot p(\gamma) g^{\mu \nu},
\ee
\ie, the tensor structure that arises from operator (\ref{eq:vectordecayop}).

\subsection{Dilepton versus $h^0 \gamma $}

The form of the operator responsible for the decay $Z' \rightarrow \gammah$ immediately indicates that a vector resonance decaying to $\gammah$ is likely to exhibit decays to lepton pairs.
Specifically, as can be seen by replacing $h$ with its VEV, whatever physics gives rise to operator  (\ref{eq:vectordecayop}) should also generate a kinetic mixing of $Z'$ with the photon. A kinetic mixing of the
$Z'$ and $Z$ bosons is also likely to be present. The mixing terms in the Lagrangian can be written as 
\be
\frac{ e \, v^2}{2 (4\pi)^2 \, m_0^2} \, Z'_{\mu \nu} \left( \tilde c_\gamma \, F^{\mu \nu}  +   \frac{ \tilde c_Z}{s_W c_W} \, Z^{\mu \nu} \right) ~~.
\ee
The coefficient  $\tilde c_\gamma$ is different than $c_\gamma$ in order to take into account contributions  to the kinetic mixing which are 
 not related to electroweak symmetry breaking. The  dimensionless parameter $\tilde c_Z$ is also model dependent; $g = e/s_W$ is the $SU(2)_W$ coupling, and $c_W \equiv \cos \theta_W$,  
 $s_W \equiv \sin \theta_W$, where $\theta_W$ is the weak mixing angle.

The kinetic mixing would generically allow $Z'$ to decay to additional states, notably lepton pairs. 
The impressive sensitivity exhibited by dilepton resonance searches at the LHC means that if the dilepton and $\gammah$ decay rates are comparable, $\Gamma(Z' \to \ell^+ \ell^-) \gsim \Gamma(Z' \to \gammah)$, the new resonance may be first observed in dileptons.
In this case, an $\gammah$ search would be an important part of fully characterizing the $Z'$, and as a probe of the physics responsible for generating the kinetic mixing and loop-level decay $Z' \to \gammah$.
The $\gammah$ search would be facilitated by knowing $M_{Z'}$ from the dilepton search.
Alternatively, if the dilepton decay rate is subdominant, $\gammah$ may represent a viable discovery channel.

To elucidate which situation may be most likely in different regions of parameter space, let us estimate the relative rates of the $Z' \to \gammah$ and $Z' \to \ell^+\ell^-$ channels.
The partial widths are 
\be
\Gamma (Z' \to \gammah) =   \frac{c^2_\gamma \,  \alpha \, v^2}{ 1536 \pi^4 \, m_0^4} \,  M_{Z'}^3 \left(1 - \frac{M_h^2}{M_{Z'}^2}\right)^{\! 3}  ~~,
\ee 
and, to leading order in the parameters $\tilde c_\gamma, \tilde c_Z$ (\ie, assuming the kinetic mixing is small)
\be
\Gamma (Z' \to e^+e^-) = \Gamma (Z' \to \mu^+\mu^-) =  
 \left( \tilde c_\gamma ^2 + \tilde c_\gamma \tilde c_Z  a_W 
 + \tilde c_Z^2  b_W \right) \frac{ \alpha^2  v^4}{192 \pi^3 \, m_0^4}  \, M_{Z'}  ~~,
\ee 
where  we defined
\bear
&& a_W = \frac{ 1 - 4 s_W^2}{2 c_W^2s_W^2} + O(M_Z^2/M_{Z'}^2) \approx 0.21  ~~~,
\nonumber \\ [2mm]
&& b_W = \frac{ 1 - 4 s_W^2 + 8 s_W^4}{8 c_W^4 s_W^4} + O(M_Z^2/M_{Z'}^2)    \approx  2.0   ~~~.
\eear
Thus, the ratio of branching fractions to $\gammah$ and lepton pairs ({\it i.e., } the sum over $e^+e^-$ and $\mu^+\mu^-$)
can be parametrized as follows:
\be
\frac{B (Z' \to \gammah )  }{B (Z' \to  \ell^+\ell^-) } =  r_{h\gamma} \,  \left(\frac{M_{Z'}}{1 \; {\rm TeV}} \right)^{\! 2}  
\left(1 - \frac{M_h^2}{M_{Z'}^2}\right)^{\! 3}  ~~,
\label{eq:ratio}
\ee
where $r_{h\gamma}$ is a dimensionless parameter that depends on the coefficients $c_\gamma$, $\tilde c_\gamma$ and $\tilde c_Z$.

Consider the case where the only $SU(2)_W \times U(1)_Y$-invariant operator responsible for $Z' \to \gammah$ and $Z' \to \ell^+\ell^-$ is 
\be
\frac{C_H \, e}{(4\pi)^2 \, m_0^2 \, c_W} \, H^\dagger H \,  Z'_{\mu \nu} B^{\mu \nu}  ~~,
\label{eq:singletoperator}
\ee
where $H$ is the SM Higgs doublet, and $B^{\mu\nu}$ is the hypercharge field strength.
The coefficients then satisfy  $c_\gamma = \tilde c_\gamma = C_H $  and $\tilde c_Z = - C_H  s_W^2$, so that $r_{h\gamma} \approx 40$.
From Eq.~(\ref{eq:ratio}) it then follows that $B (Z' \to \gammah ) > B (Z' \to  \ell^+\ell^-)$ for $M_{Z'} \gtrsim 247$ GeV.
The above operator also induces a decay to $h^0 Z$, which we do not discuss here as it is a less sensitive search mode than $h^0\gamma$
due to the small leptonic $Z$ branching fraction.

Another example of an operator that induces $Z'$ decays into $\gammah$ and $\ell^+\ell^-$  is
\be
- \frac{C'_H \, g}{(4\pi)^2 \, m_0^2 } \, H^\dagger  \sigma^a  
 H \,  Z'_{\mu \nu} W^{a \mu \nu}  ~~.
 \label{eq:tripletoperator}
\ee
If this single operator contributes to these decays, then $c_\gamma = \tilde c_\gamma = C'_H  $ and $\tilde c_Z = C'_H  c_W^2$, which implies 
 $r_{h\gamma} \approx 18$, and the $ \gammah$ branching fraction is larger than the dilepton one for $M_{Z'} \gtrsim 309$ GeV.

The coefficients of the kinetic terms, $\tilde c_\gamma$ and $\tilde c_Z$, can also receive contributions which are independent of electroweak symmetry breaking.  
We assume that the $Z'$ boson is associated with a  $U(1)$ gauge symmetry.
A tree-level dimension-4 operator $Z'_{\mu \nu} B^{\mu \nu}$ may be eliminated by embedding one of the Abelian gauge groups in a larger group at some high scale.
If there are fields that carry both hypercharge and $U(1)$ charges, though, this kinetic mixing may be generated at one loop, but with a model dependent coefficient. 
In the particular case where all the fields charged under both groups have the same mass and the product of hypercharge and $U(1)$ 
charge summed over all fields is zero, the 1-loop contribution to $Z'_{\mu \nu} B^{\mu \nu}$ vanishes. Thus, it is possible that the dominant contribution to 
$Z' \to \ell^+\ell^-$ arises from  the operators  (\ref{eq:singletoperator}) or (\ref{eq:tripletoperator}). This might not be the case 
if the SM quarks carry the new $U(1)$ charges, as explained towards the end of the next subsection.
 
\subsection{Dijet versus $h^0 \gamma $}

Kinetic mixing also generates a decay to dijets. The large background to dijet resonance searches ensures that, for $\Gamma(Z' \to jj)$ comparable to $\Gamma(Z' \to \gammah)$ or $\Gamma(Z' \to \ell^+ \ell^-)$, $\gammah$ and dilepton searches would represent more promising discovery channels for the new state. However, a model with an appreciable $Z'$ production (and hence $\gammah$) rate typically implies a significant decay rate to dijets, beyond the small rate induced via kinetic mixing. In particular, the $Z'$ production at the LHC is large provided there are tree-level couplings of the first-generation quarks to the $Z'$.  In that case the width of the $Z'$ decay mode into a pair of jets is several  orders of magnitude larger than the width of $Z' \to h^0 \gamma $. Nevertheless, due to the large dijet background, $h^0 \gamma $ could be the discovery mode 
in some cases. 
In this subsection, we consider a model based on a leptophobic $Z'$, and demonstrate that an interesting rate for $pp \to Z' \to \gammah$ can still be achieved consistent with constraints from dijet resonance searches, particularly at lower $M_{Z'}$.

\begin{figure}[t] 
   \centering
 \includegraphics[width=0.72\textwidth, angle=0]{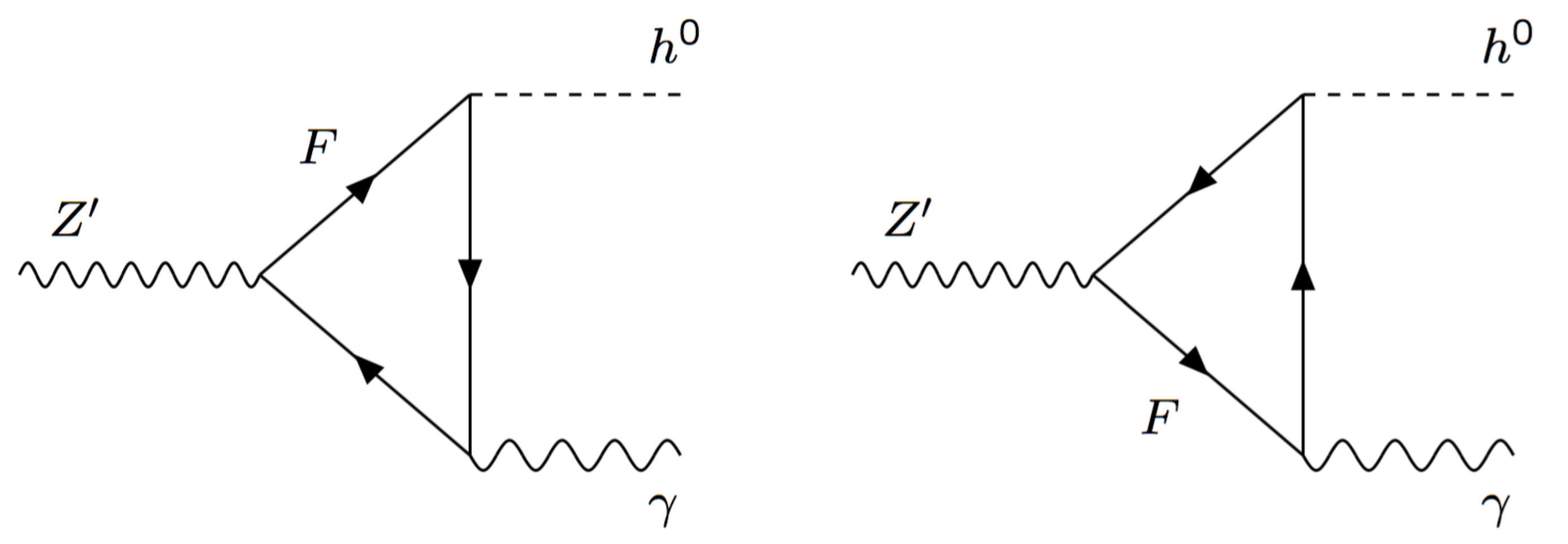}
   \caption{One-loop contributions of a fermion $F$ to the $Z' \to h^0 \gamma $ amplitude.}
   \label{fig:1loopdiagram}
\end{figure}

Let us compute precisely the 1-loop $h^0 \gamma $ width,   $\Gamma(Z'\rightarrow \gammah)$, in the case where 
a fermion $F$ of mass $m_F$ and electric charge $Q$ has a Yukawa coupling $(y_F/\sqrt{2}) \,  h^0 \bar F F $ to the Higgs boson, and 
couples vectorially to the $Z'$:
\be
\frac{g_z}{2}  z_F \; Z'_\mu   \,  \left( \overline F_L \gamma^\mu F_L +  \overline F_R \gamma^\mu F_R \right)  ~~~,
\ee
where $g_z$ is the gauge coupling, and $z_F$ is the charge  of the fermion under the new gauge group.
There are two diagrams, shown in  Figure~\ref{fig:1loopdiagram},
which contribute to $Z' \to h^0 \gamma $. 
Defining the mass ratios 
\be
r_h = \frac{M_h}{M_{Z'}}     \;\;\; , \;\;\;      r_F = \frac{m_F}{M_{Z'} }  ~~~,
\ee
we find that the 1-loop width  induced by fermion $F$ is 
\be
\Gamma(Z'\rightarrow \gammah) = \frac{( N_c Q_F )^2 \alpha}{384 \pi^4} ( y_F g_z z_F)^2  
\, M_{Z'}  \, f(r_h^2, r_F^2)    ~~,
\label{eq:loopwidth}
\ee
where $N_c$ is the number of colors of the fermion. The dimensionless function $f$, computed in the Appendix, 
includes the loop integral and the phase space.

\begin{figure}[t]
\begin{center}
\includegraphics[width=0.495\textwidth]{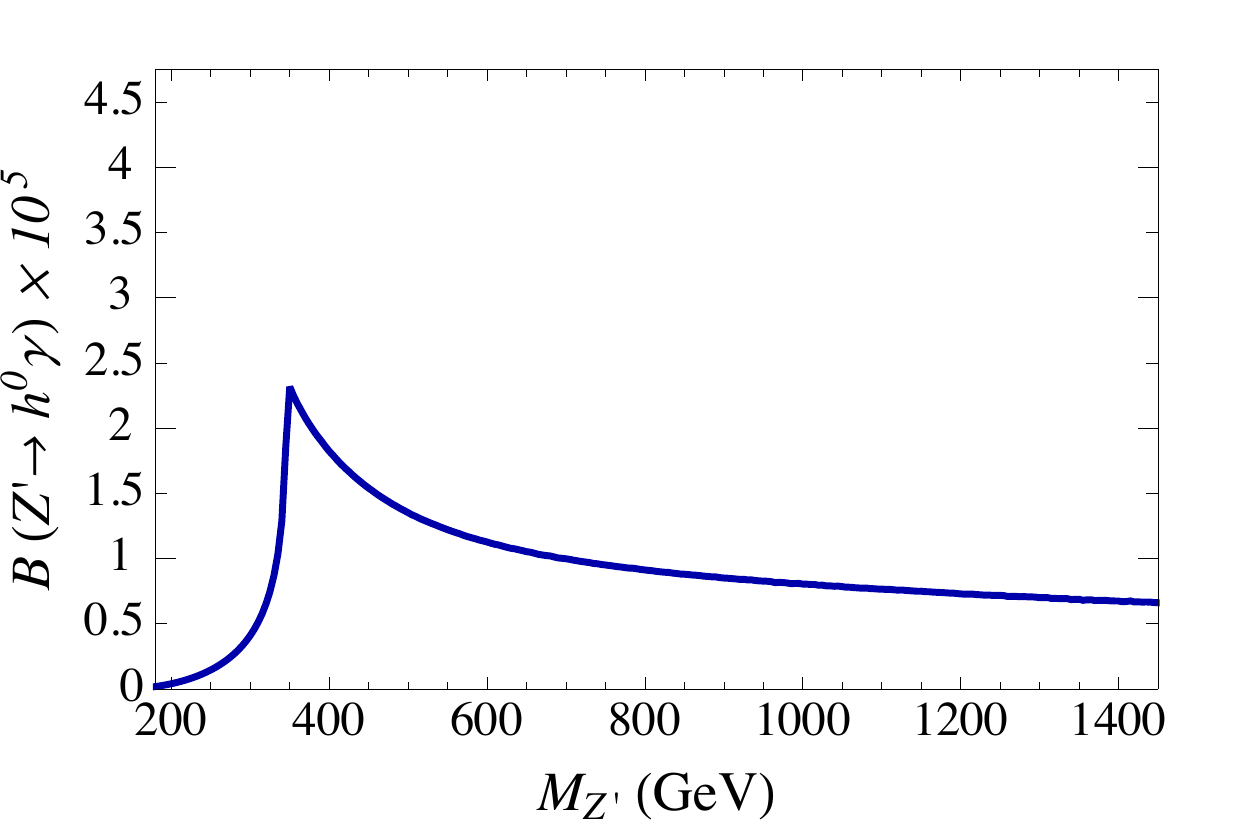} \hspace*{-4mm} \includegraphics[width=0.5\textwidth]{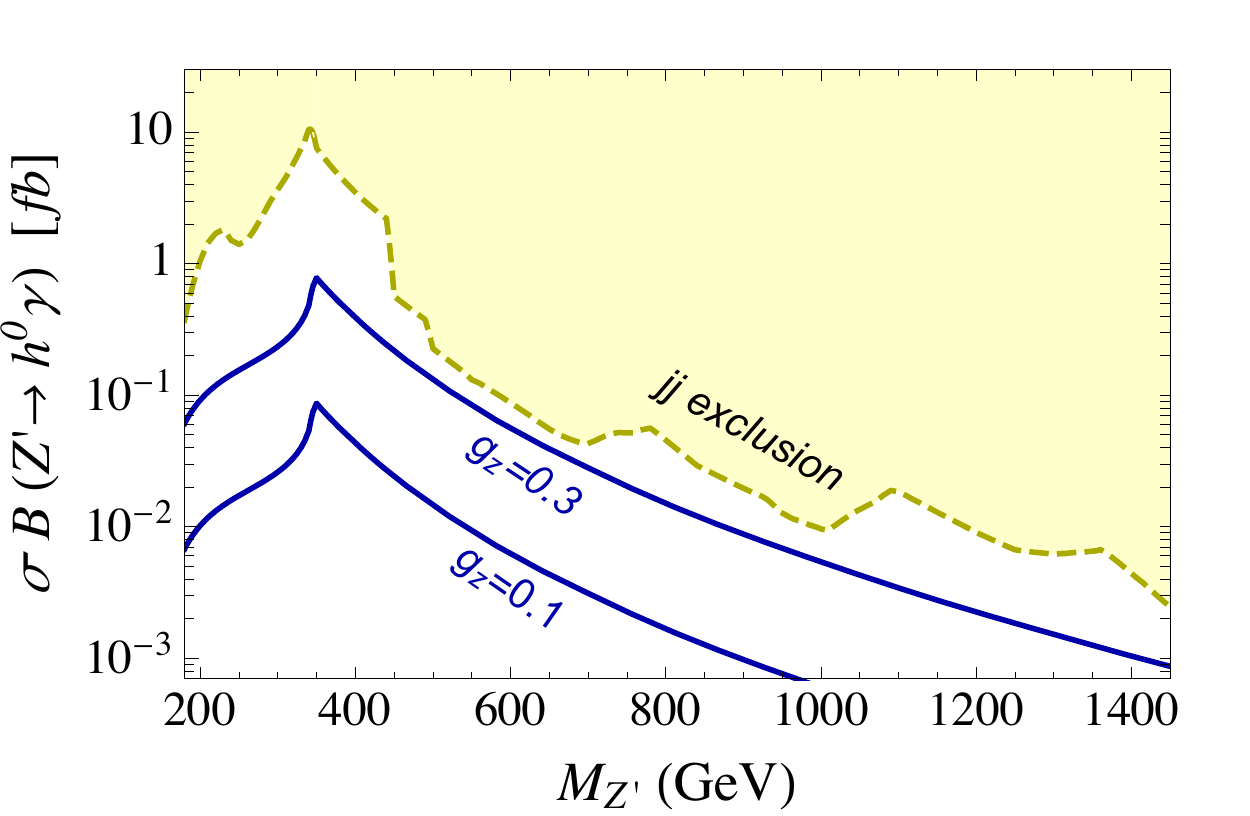} 
\caption{{\it Left panel:} The $Z'\rightarrow \gammah$ branching fraction as a function of the $Z'$ mass, when the $Z'$ has  flavor-universal vector couplings to the SM quarks, 
and there are no new particles running in the loop.
{\it Right panel:} Leading-order $pp \to Z'\rightarrow \gammah$ cross section at $\sqrt{s} =  13$ TeV 
for two values of the gauge couplings $g_z$, when all the SM quarks have charge $z_F=1/3$ under the new gauge group. The shaded region is ruled out by current dijet resonance limits.
}
\label{fig:BRplot}
\end{center}
\end{figure}

In the case where $F$ is the top quark, $N_c = 3$, $Q_F = 2/3$, and  $y_F$ is the SM top Yukawa coupling: $ y_t =  \sqrt{2} \, m_t /v \approx 1$. Note that the contributions
from lighter quarks are suppressed by their mass squared.
Assuming flavor-universal vector couplings of $Z'$ to the SM quarks, and that no other particles contribute to the loop process,
the ratio of the $ \gammah$ and dijet widths is
\be
\frac{\Gamma(Z'\rightarrow \gammah)}{ \sum_{q} \Gamma(Z'\rightarrow q\bar{q} ) }  \simeq 
\frac{ \alpha \,  y_t^2 \, f(r_h^2, r_t^2)}{  6 \pi^3 \left( 5+ \left(1- r_t^2\right)\sqrt{1-  4 r_t^2}  \right) } ~~,
\label{ref:widthratio}
\ee
where $r_t \equiv m_t/M_{Z'}$.
If the $Z'$ does not interact with new particles coupled to the Higgs doublet,
then the maximum value of the $Z'\rightarrow \gammah$ branching fraction 
occurs for $M_{Z'} = 2 m_t$: 
$ B(Z' \to h^0\gamma)^{\rm max}  = 2.3 \times 10^{-5}$.
The $\gammah$ branching fraction is plotted as a function of $M_{Z'}$  in the left panel of Figure~\ref{fig:BRplot}.

A $Z'$ boson with flavor-universal couplings to all SM quarks arises in the presence of an
extension of the SM gauge group by a $U(1)_B$ symmetry, with all quarks carrying the same charge (by convention $z_F = 1/3$ while the 
gauge coupling $g_z$ is a free parameter).  
The cancellation of the gauge anomalies involving $U(1)_B$ requires new fermions (called anomalons), which must be 
chiral with respect to $U(1)_B$,  and are constrained to be vectorlike with respect 
to the SM gauge group. Specific sets of anomalons were introduced in Refs.~\cite{Carena:2004xs,Dobrescu:2013coa,Dobrescu:2014fca}.
The couplings of the anomalons to the Higgs doublet are model dependent. In the limit where these vanish, the anomalons do not contribute to the $Z' \to \gammah$ width. 

The $Z'$ production cross section is proportional to $g_z^2$.
We have computed the inclusive leading-order production cross section of the $Z'$, $\sigma(pp \to Z' + X)$,  at the 13 TeV LHC using
MadGraph \cite{Alwall:2014hca}, with model files generated by FeynRules \cite{Alloul:2013bka}.
In the right panel of Figure~\ref{fig:BRplot} we plot the cross section times the 
branching fraction,  $\sigma(pp \to Z' + X) B(Z' \to \gammah)$,  as a function of $M_{Z'}$, for $Z'$  gauge coupling $g_z = 0.3$ or 0.1. 
We also show the upper limit imposed by various dijet resonance searches \cite{CMS:2017xrr}, which constrain the gauge 
coupling \cite{Dobrescu:2013coa} in this leptophobic $Z'$ model. 
Even though the loop generated decay has a small branching fraction, 
the $h^0\gamma$ resonance searches can still compete with the dijet resonance searches.
The $\sigma(pp \to Z' \to \gammah)$ cross section at $\sqrt{s} = 13$ TeV can be larger than 1 fb for $M_{Z'} < 450 $ GeV, 
while values larger than 0.1 fb are allowed for $M_{Z'} < 550 $ GeV.

If particles beyond the SM carry electric and $U(1)_B$ charges, and also couple to the Higgs boson, then 
their 1-loop contributions interfere with the SM quark loops and may enhance or decrease the $Z' \to \gammah$ 
branching fraction. Let us consider a simple extension of the SM with two vectorlike leptons carrying $U(1)_B$ charge $-1$: 
one is a weak-doublet of hypercharge $-1/2$ (same as the SM lepton doublets) labelled $\psi_D = (\psi_D^\nu , \psi_D^e)$, 
and the other one is a weak-singlet of hypercharge $-1$ labelled $\psi_S$. These have gauge-invariant masses
as well as a Yukawa coupling to the SM Higgs doublet,
\be
- m_D \bar \psi_D \psi_D -  m_S \bar \psi_S \psi_S - y_\psi \left( \bar \psi_D \psi_S H  + {\rm H.c.} \right)  ~.
\ee
The two electrically-charged fermions mix, giving rise to the following mass-eigenstates:
\be
\left( \ba{c} \psi_1 \\ \psi_2 \ea \right) = \left( \ba{cc} \cos\theta & \sin\theta \\ - \sin\theta & \cos\theta \ea \right) \left( \ba{c} \psi_D^e \\  \psi_S \ea \right)   ~~,
\ee
where the mixing angle satisfies
\be
\tan 2\theta = \frac{\sqrt{2} y_\psi v }{m_D - m_S}   ~~.
\ee
Labelling the mass of the lightest physical state by $m_\psi$, the mass of the other charged vectorlike lepton is 
\be
m_{\psi'} = m_\psi + \frac{\sqrt{2}\, y_\psi v }{\sin 2\theta}  ~~.
\ee

 \begin{figure}[t]
\begin{center}
\includegraphics[width=0.495\textwidth]{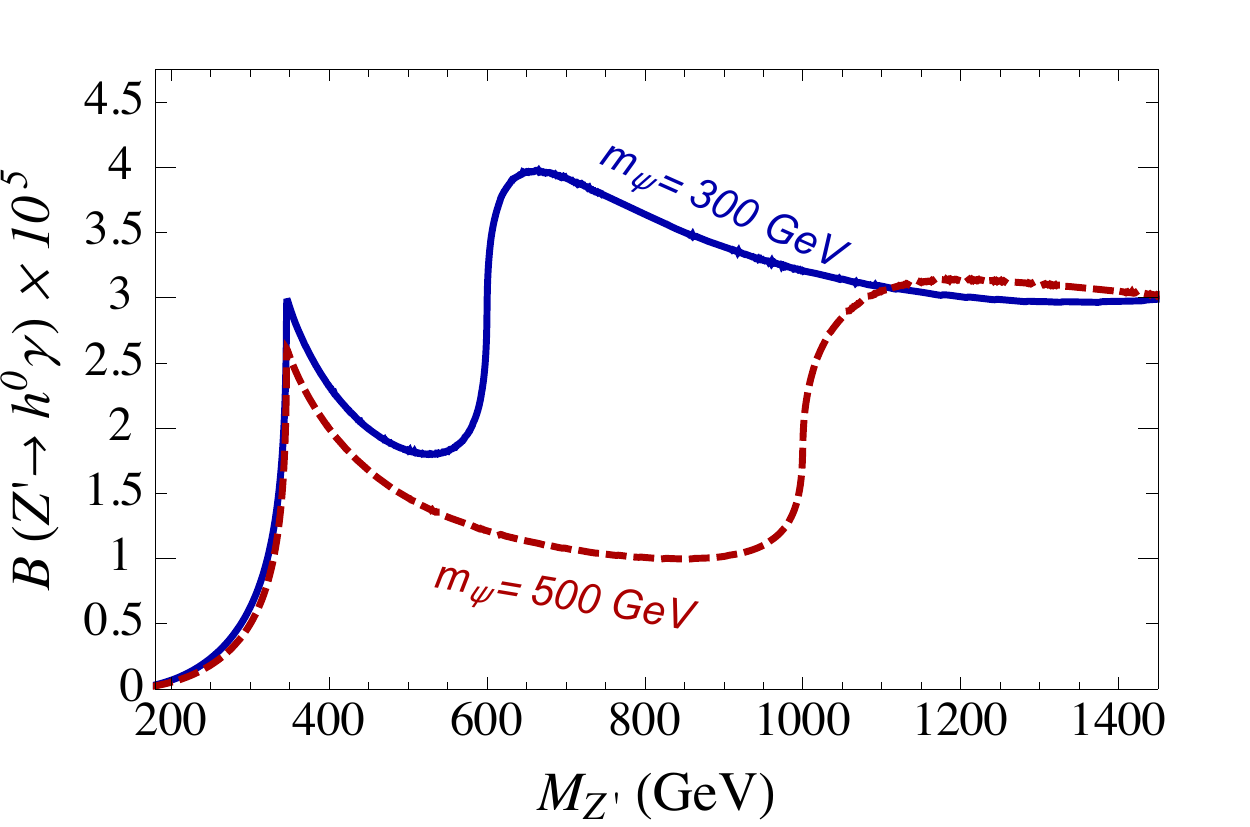} \hspace*{-4mm} \includegraphics[width=0.5\textwidth]{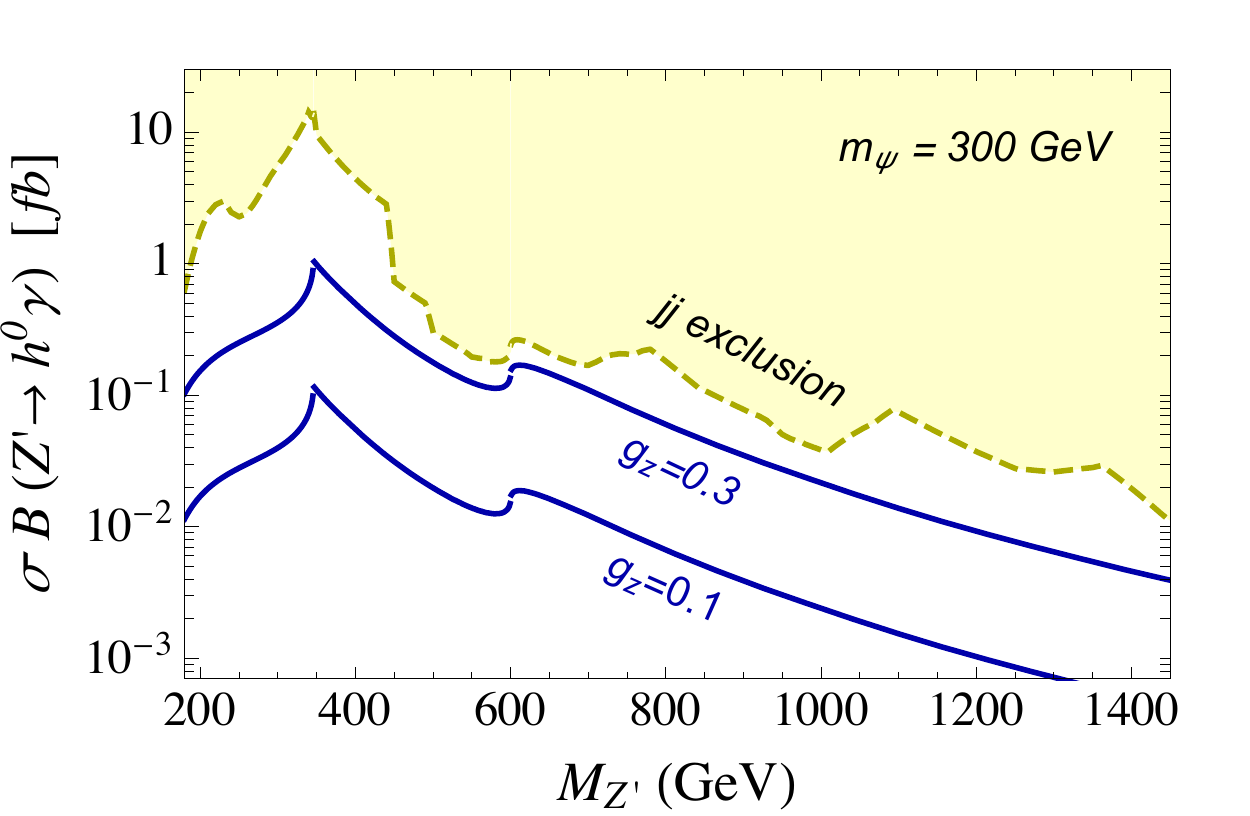} 
\caption{{\it Left panel:} The $Z'\rightarrow \gammah$ branching fraction for flavor-universal vector couplings of  $Z'$ to the SM quarks with $z_F = 1/3$,
when two vectorlike leptons with $z_\psi = -1$ are running in the loop. The mass of lightest vectorlike lepton is $m_\psi = 300$ GeV (blue solid line)  or 500 GeV (dashed red line); the
mixing angle and the Higgs Yukawa coupling of the vectorlike leptons are fixed at $\sin\theta = 0.3$ and $y_\psi = 1$, respectively.
{\it Right panel:} Leading-order $pp \to Z'\rightarrow \gammah$ cross section at $\sqrt{s} =  13$ TeV 
for $m_\psi = 300$ GeV and $g_z = 0.3$ or 0.1. The shaded region is excluded by dijet resonance searches.
}
\label{fig:BRplotVL}
\end{center}
\end{figure}

The $Z' \to \gammah$ width,  given in Eq.~(\ref{ref:widthratio}) when the only large contribution is from the top quark, is modified in this case by a factor of
\be
\left| 1 +  \frac{3 y_\psi }{2 y_t} \, \sin 2\theta \, \frac{ I \!\left( r_h, r_\psi \right) + I \!\left( r_h, r_{\psi'} \right)}{I \!\left( r_h, r_t \right)  }   \right|^2  ~~,
\ee
where $r_\psi = m_\psi/M_{Z'}$, $r_{\psi'} = m_{\psi'}/M_{Z'}$, and $I$ is the loop integral given in Eq.~(\ref{eq:FPintegral}).
The $Z' \to \gammah$ branching fraction is shown in Figure \ref{fig:BRplotVL} for $\sin\theta = 0.3$, $y_\psi = 1$ and $m_\psi = 300$ GeV or 500 GeV. 
The $pp \to Z' \to \gammah$ cross section at the 13 TeV LHC is shown in the right panel of Figure \ref{fig:BRplotVL} for two values of the gauge coupling.

The limits on vectorlike lepton masses from collider searches are model dependent, and are rather loose for decays 
into $W\nu$, $\tau Z$, or $\tau h^0$ \cite{Falkowski:2013jya}. When the vectorlike leptons are heavier than the $Z'$ boson, they can decay 
into a $Z'$ and a SM lepton, but again the LHC sensitivity is reduced when the lepton is a $\tau$.  
The constraints on the Higgs Yukawa coupling of the vectorlike leptons from 
the measurements of $h^0 \to \gamma \gamma$ are also loose.

Let us now comment on the expected size of the kinetic mixing and associated dilepton decay rate in this model.
To ensure $Z' \to \ell^+ \ell^-$ does not dominate over $Z' \to \gammah$, the tree-level kinetic mixing between the $U(1)_Y$ and $U(1)_B$ gauge bosons is assumed to vanish at a scale $\Lambda$, of order 10 TeV or higher. As mentioned earlier, this can be enforced, for example, by embedding one of the $U(1)$ gauge groups in a non-Abelian group at scale $\Lambda$.
Furthermore, for the minimal set of anomalons given in  \cite{Dobrescu:2014fca}  the sum over anomalons of the product of hypercharge and $U(1)_B$ charge cancels the analogous sum over SM quarks.
As the field content of this model satisfies Tr$(YB) = 0$, the dimension-4 kinetic mixing is not induced at one loop above the mass of the anomalons, $m_a$.
Below this mass, however, the $Z_{\mu\nu}' B^{\mu\nu}$ kinetic mixing parameter will run.  Since the contribution from the anomalons cancels against the contribution from the SM fermions, the ratio of the 
kinetic mixing to $c_\gamma$ is 
proportional to $\log(m_a/M_{Z'})$.  Thus, if the anomalon and $Z'$ masses are comparable, then  $\tilde c_\gamma$ and  $\tilde c_Z$ are of the same order as  $c_\gamma$. As a result, $\gammah$ and dilepton resonance searches represent complementary probes of the model, and either may serve as a possible discovery channel.

While the couplings of anomalons to the SM Higgs boson are model dependent and may even vanish, the anomalon couplings to the scalar field, $\phi$, that breaks $U(1)_B$ are necessary to generate their masses.  Thus, anomalon loops induce an additional dimension-6 operator:
\be
\frac{C_\phi \, e}{(4\pi)^2 \, m_0^2 \, c_W} \, \phi^\dagger \phi \,  Z'_{\mu \nu} B^{\mu \nu}  ~~.
\label{eq:phioperator}
\ee
This contributes to the kinetic mixing when $\phi$ is replaced by its VEV, but also leads to other notable experimental signatures.
The SM Higgs boson and the CP-even component of  $\phi$ can mix through the Higgs portal $|\phi|^2 |H|^2$. The physical states are the observed Higgs-like particle ($h$) of mass $M_h \approx  125$ GeV, and a 
second scalar $h'$ of mass $M_{h'}$. The mixing leads to $h'$ decay to pairs of SM fermions and vector bosons. For $M_{h'} < M_{Z'}$, the decay $Z'\rightarrow h' \gamma$ with $h'$ decaying as a
Higgs-like particle could also be searched for.  For $10\,\GeV \lsim M_{h'} \lsim 160\,\GeV$ the search channel would be $b\bar{b}\gamma$ with the $b\bar{b}$ resonance no longer at $M_h $.  For higher $M_{h'}$
masses a combination of $W^+W^-\gamma$, $ZZ\gamma$, and $t\bar{t}\gamma$, again with subresonances at the $h'$ mass, would be the dominant search channels.

\section{Alternative models with a Higgs-photon resonance}
\label{sec:altmodels}
\setcounter{equation}{0}

In the preceding section, we discussed models that include a spin-1 particle that could give rise to an $\gammah$ signal at the LHC. 
While this motivates the implementation of a dedicated $\gammah$ resonance search, 
$Z' \rightarrow \gammah$ seems to be most interesting for $M_{Z'} \lsim 800 \GeV$, where the weaker dijet limits still permit the subdominant, loop-induced
 $\gammah$ decay to be sufficiently large to be observed at the LHC.
In this section we briefly discuss other models that may be explored by an $\gammah$ search, and may represent candidates for new physics should a signal be observed in other regions of parameter space. 

 One challenge  for the decay $Z' \rightarrow \gammah$ in the models discussed so far is the small branching fraction, of order $10^{-5}$ in the presence of tree-level couplings to quarks, which are necessarily present if the $Z'$ is directly produced in proton-proton collisions.
An alternative is that the $Z'$ boson is not produced directly but rather in a cascade of some heavier particle, $X_{\rm heavy}$, for instance in association with other jets or particles that escape the detector as missing energy. In such a model, the $Z'$ couplings to quarks may vanish at tree-level, and so the decay to $\gammah$ may have a large branching fraction. Moreover, if the additional jets are soft or the missing energy is small, the signal may still appear similar to that of a directly-produced resonance. A similar scenario was considered in the context of diphoton resonances in \cite{Dobrescu:2016owc}. There, it was noted that a loop-induced diboson decay width may still be small compared to a 3-body decay through an off-shell $X_{\rm heavy}$ to jets (perhaps plus missing energy), particularly if the splitting between $X_{\rm heavy}$ and the diboson resonance is small. A similar caveat applies here.

Another way in which new physics may give rise to a large signal in an $\gammah$ search relative to that in dijets is to replace the photon by a pair of collimated photons arising from the decay of a highly-boosted light particle. 
Let us construct a model of this type, which involves two complex scalars, $\phi$ and $\phi'$, carrying the same charge  ($z_\phi$)  under a $U(1)$ gauge symmetry. 
These scalars have VEVs, so that the associated gauge boson, $Z'$, acquires a mass.
In the presence of terms such as $\phi^\dagger \phi' H^\dagger H$ in the scalar potential, the SM Higgs boson mixes with the CP-even components of $\phi$ and $\phi'$.
The CP-odd components of $H$, $\phi$ and $\phi'$ also mix, with two linear combinations becoming the longitudinal $Z$ and $Z'$ bosons; the third one
remains as a physical CP-odd scalar, $A^0$.
As a result of mixing, there is a coupling of $A^0$ to the Higgs boson $h^0$ and the $Z'$:
\be
\frac{z_\phi \, g_z}{2}  s_h \,  Z'_\mu \left( A \, \partial_\mu h  +  h \, \partial_\mu A \right)  ~~,
\ee
where $g_z$ is the gauge coupling, and  
$s_h < 1$ is a mixing parameter. This leads to a tree-level decay $Z' \rightarrow h^0 A^0$.

If $A^0$ couples at one loop to a pair of photons and its mass is below a few GeV, 
then $A^0$ will be highly boosted and the two photons will appear as a single photon in the detector. Thus, 
the $Z'$ may initially appear as a $\gammah$ resonance.
If the SM quarks are charged under the $U(1)$ gauge group, then the 
$Z'$ gauge boson is produced at tree-level at the LHC, and it can also decay into a SM quark-antiquark pair.
In the case of flavor-universal quark charges, equal to 1/3 as in Section 3.2, the ratio of the $Z'$  widths into $h^0 A^0$  and quark pairs is 
\be
\frac{\Gamma(Z'\rightarrow h^0 A^0)}{ \sum_{q} \Gamma(Z'\rightarrow q\bar{q} ) }  \simeq 
\frac{ 3 z_\phi^2 \, s_h^2 \, (1- r_h^2)^3}{  2 \left( 5+ \left(1- r_t^2\right)\sqrt{1-  4 r_t^2}  \right) } ~~,
\label{ref:widthratioA0}
\ee
where again $r_h = M_h/M_{Z'}$ and  $r_t = m_t/M_{Z'}$. 
For large $M_{Z'}$, the branching fraction is $B(Z'\rightarrow h^0 A^0) \approx s_h^2 z_\phi^2 /4$.
Measurements of the Higgs boson \cite{Khachatryan:2016vau} currently constrain $s_h^2 \lsim 0.2$, so that $B(Z'\rightarrow h^0 A^0)$ may be as large as 5\% for $z_\phi = 1$.

The coupling of $A^0$ to photons is induced at one loop. The particles running in the loop may, for example, be vectorlike leptons of electric charge $Q_L$ and mass $m_L$.
There is also a 1-loop coupling of $A^0$ to gluons, induced by a top quark loop, which is suppressed by a mixing parameter, $s_A$, from the CP-odd sector.
The ratio of the $A^0$ widths into photons and gluons may be approximated for $M_A$ above a GeV by treating the gluons as massless jets:
\be
\frac{\Gamma(A^0\rightarrow \gamma\gamma )}{ \Gamma(A^0 \rightarrow gg) }  \simeq  \frac{\alpha^2}{2 \alpha_s^2}
\left(  Q_L^2  \,  \frac{y_{LA} \, m_t}{ y_{tA} \, s_A \, m_L} + \frac{4}{3} \right)^{\, 2}  ~~,
\ee
where $y_{LA}$ and  $y_{tA}$ are the Yukawa couplings of $A^0$ to the vectorlike lepton and the top quark, respectively. We neglected here the higher-order corrections in the mixing parameters.
Let us choose a benchmark point in the parameter space: $Q_L =2$, $y_{LA} = y_{tA}$, $m_L = 400$ GeV, $s_A = 0.3$. 
For these values, the $A^0\rightarrow \gamma\gamma$ branching fraction is then above 20\%.

When the branching fraction of $A^0 \to \gamma\gamma$ is large, this leads to values at the percent level for an effective 
$B(Z'\rightarrow h^0 ``\gamma ")$, where $``\gamma "$ stands for a pair of collimated photons. This is an increase by three orders of magnitude compared to the $\gammah$ branching fraction obtained in Section 3, potentially turning the $\gammah$ resonance search into a discovery channel.

The production cross section of the $Z'$ at $\sqrt{s} = 13$ TeV decreases from 100 pb at $M_{Z'} = 300$ GeV to 60 fb at $M_{Z'} = 2$ TeV, for $g_z = 0.4$, which is approximately the largest gauge coupling 
allowed by current dijet resonance searches \cite{Dobrescu:2013coa}. Thus,  the 
cross section times branching fraction may be as large as 0.6 fb for a mass up to 2 TeV. Using the limit projection shown in Figure 1, we conclude that an $h^0``\gamma"$ resonance search 
for a  2 TeV $Z'$ may set a limit with 300 fb$^{-1}$ of data, or may lead to discovery with 3000 fb$^{-1}$.

In any model featuring a decay to $``\gamma "$, the parameter space will be limited by the requirement that the light state $A^0$ decays to two photons before reaching the electromagnetic calorimeter and is indeed reconstructed as a single photon \cite{Dobrescu:2000jt}.  Although the diphoton decay can have a sizable branching fraction, the lifetime is dominated, for masses above a GeV, by the width into gluons,
\be
\Gamma(A^0 \rightarrow gg) \simeq \frac{\alpha_s^2 \, y_{tA}^2 s_A^2 }{64\pi^3 \, m_t^2 } M_A^3~~.
\ee
The typical separation between the photons produced in the decay of a state with a total  width $\Gamma_A$ and boost $\gamma_A$, at distance $R$ from its production, is
\be\label{eq:photonangularseparation}
\Delta \theta_{\gamma\gamma} \approx \left(1-\frac{\gamma_A \beta_A}{R\Gamma_A}\right)\frac{2}{\gamma_A}~.
\ee
In order that the collimated photons are reconstructed as a single photon, this opening angle cannot be too large. For instance, the innermost calorimeter layer at ATLAS is designed to reject pions and has segmentation $\Delta \eta_{\rm ATLAS} = 3 \times 10^{-3}$, 
placing an upper bound on $M_A$ that grows with the $Z'$ mass (due to the $A^0$ boost).  Meanwhile, the $A^0$ must decay on average before the inner layer of the calorimeter, a distance of $O(1.3)$ m, which places a lower bound on $M_A$.  While it is possible to simultaneously satisfy these constraints, a full analysis should take into account the temporal and angular distributions of $A^0$ decays.

So, for certain choices of $M_A$ and $M_{Z'}$, it may be possible that a resonance be observed in an $\gammah$ search with the ``photon" comprised of collimated photons.
Subsequent studies may be able to use photon conversions or shower profile to distinguish between a single photon and boosted 
diphoton~\cite{Aad:2015bua, Agrawal:2015dbf}.
In particular, these studies may be particularly valuable if an observation is made without either a corresponding observation in dijets or dileptons, potentially suggesting these channels are forbidden or suppressed relative to the values expected in the models discussed in Section 3.

We have been discussing a new vector decaying into $h^0 ``\gamma "$ but the resonance could also be a (pseudo)scalar.  A scalar cannot decay directly to $\gammah$, but again it may appear in this channel if it decays to a pair of collimated photons, \eg, $A' \rightarrow A h^0$. Such a decay can occur via trilinear terms in the scalar potential. Moreover, if the $A'$ were produced via gluon fusion through a loop of new heavy, colored particles, the tree-level decay to a Higgs and a $``\gamma "$ could readily dominate over the loop-level decay to dijets. In this case, an $\gammah$ search may even represent the most promising approach for discovering the new resonance.


\section{Conclusions}
\setcounter{equation}{0}

The LHC experiments have carried out searches for many diboson resonances, with a notable exception: a search for a resonance consisting of the SM Higgs boson and a photon.  We have discussed simple models containing a vector resonance, $Z'$, that decays at one loop to this $\gammah$ final state.  The $Z'$ is produced from its coupling to light quarks so the branching fraction to $\gammah$ is of the order of 
$\sim 10^{-5}$.  Typically one would also expect the $Z'$ to decay to lepton pairs at a comparable rate (though this need not be the case), and to a pair of jets nearly 100\% of the time.  Despite the small branching fraction, there are viable models with $pp \to Z'\rightarrow \gammah$ cross section 
larger than 1 fb, for $M_{Z'}$ in the 200--450 GeV range.  

Larger branching fractions to the $\gammah$ final state can be achieved if the photon is not a single photon but rather a collimated pair of photons produced in the decay of a light pseudoscalar, $Z'\rightarrow hA^0\rightarrow h ``\gamma"$.   For resonance mass above $\sim 1$ TeV the decay products of the Higgs boson would be boosted and, for the dominant $b\bar{b}$ mode, would be reconstructed as a single jet with substructure.  Excited quark searches look for a jet+photon final state and we recast the existing searches to estimate the reach of a $q^*$ search, augmented by a Higgs tagger applied to the jet.  We estimated it is possible to discover a $Z'$ using this technique up to masses of about 2 TeV, and we presented a model where the branching fraction to $h^0 ``\gamma"$ is relatively large, at the percent level.  

Although the decay of a spin-0 particle into $h^0 \gamma$ is forbidden, a heavy neutral scalar scalar may cascade decay into the Higgs boson and collimated photons, leading again to  an $h^0 ``\gamma"$
final state. The production cross section of a scalar from gluon fusion is smaller than that of a $Z'$ coupled at tree level  to  first-generation quarks, so the mass range accessible for discovery in this case is reduced.
If a Higgs-photon resonance is discovered, the angular distribution may be needed to identify the spin of the heavy particle. 

In addition to resonances decaying to a SM Higgs and a photon it is possible that there are other Higgs-like scalars, $h'$, that could be produced in the decay  $Z' \rightarrow h' \gamma$.  For $h'$ mass below the $W^+W^-$ threshold the final states are the same as for $\gammah$ but the kinematics are different.
Dedicated analyses for $\gammah$ and $h'\gamma$, at both light and heavy resonance masses, could uncover a new boson coupled to quarks (a $Z'$, or even a scalar in the case of collimated photons), 
or allow for further characterisation of a resonance found through other final states, {\it e.g.},  lepton or  jet pairs. We strongly advocate for the inclusion of these final states in the diboson search program.

\bigskip

{\it \bf Acknowledgments:}  
This work was supported by the DoE under contract number DE-SC0007859 and Fermilab, operated by Fermi Research Alliance, LLC under contract number DE-AC02-07CH11359 with the United States Department of Energy.


\appendix
\section{Appendix: The loop integral}\label{sec:appendix}     \setcounter{equation}{0}

The contributions of a fermion loop (see Figure \ref{fig:1loopdiagram}) to the $Z' \to \gammah$ width depend on a function $f$ introduced in 
Eq.~(\ref{eq:loopwidth}). 
This is given by 
\be
 f(r_h^2, r_F^2) =  \left(1 - r_h^2 \right)^{\! 3}  \;  \left| I \!\left( r_h, r_F \right)  \right|^2   ~~~,
\ee
where the first term is a phase-space factor, and the second term arises from the loop integral.

\begin{figure}[t]
\begin{center}
\includegraphics[width=0.69\textwidth]{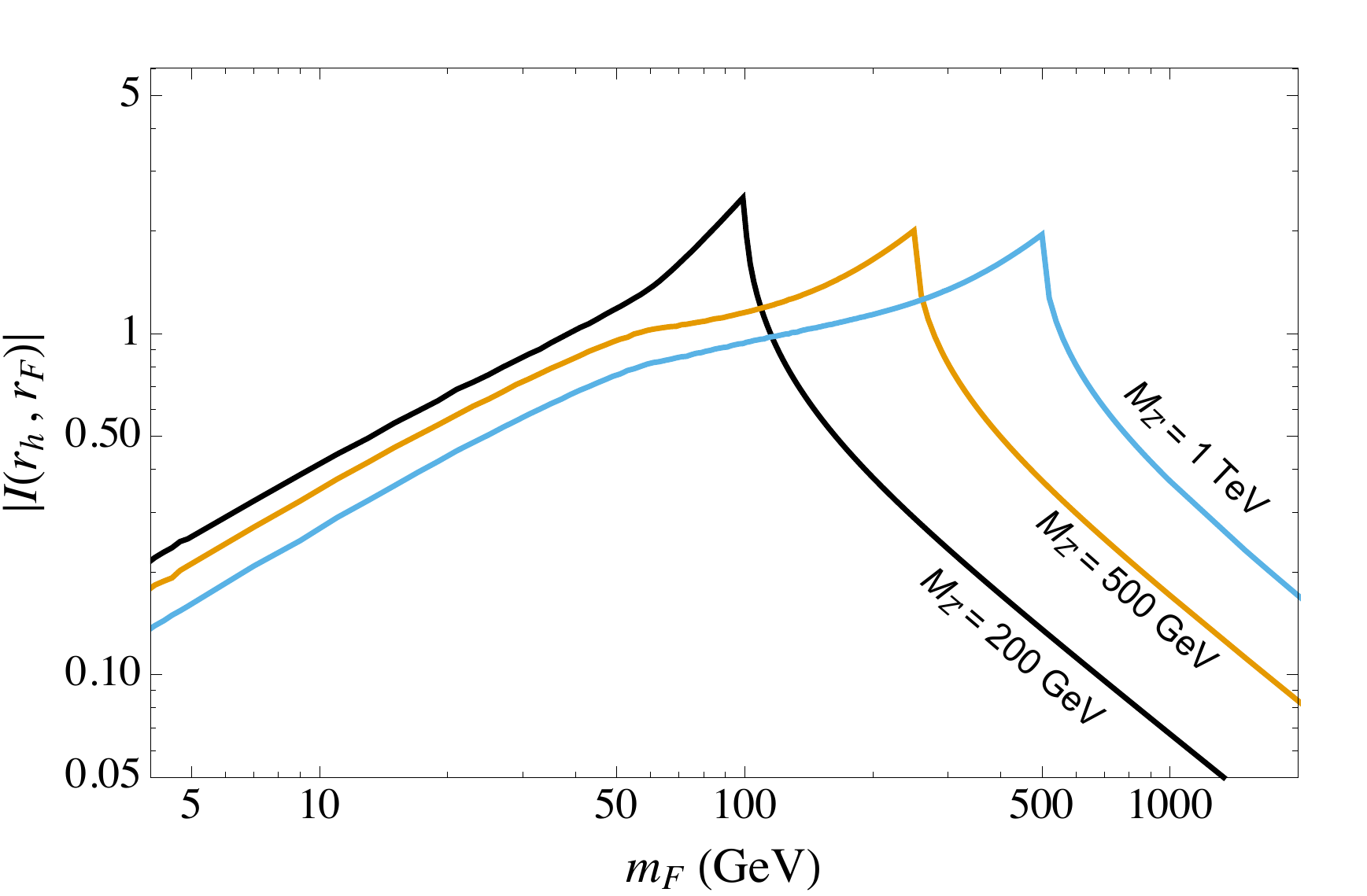} 
\caption{The loop integral $|I(M_h/M_{Z'},m_F/M_{Z'})|$ of Eq.~(\ref{eq:FPintegral}) as a function of the fermion mass, for three values of $M_{Z'}$. }
\label{fig:FPintegral}
\end{center}
\end{figure}

The two diagrams shown in Figure \ref{fig:1loopdiagram} yield 
equal CP-conserving contributions, giving a factor of 2 in the amplitude.
After introducing Feynman parameters $x$ and $y$ and computing the loop momentum integrals, we find that the 
function  $ I \! \left(r_h, r_F \right) $ that enters the  $Z' \to \gammah$ amplitude is given by 
\bea
I(r_h, r_F)  \!  \!  & \! = \! &  \! r_F \int_0^1  dx\int_0^{1-x} \! dy \,  \frac{4x y-1}{(1 - r_h^2) \, xy-x(1-x)+ r_F^2 -i \epsilon}
\nonumber
\\ [3mm]
& \! = \!  & \! \frac{ r_F}{1- r_h^2} \left[ 2 + \int_0^1\frac{dx}{x}   \;
\left(  4 \, \frac{x(1-x)-  r_F^2}{1- r_h^2} -1  \right)  
\log  \left( \frac{ r_h^2 \, x(1-x)- r_F^2+i\epsilon}{x(1-x)- r_F^2+i\epsilon} \right)  \right] ~~, 
\nonumber \\
\label{eq:FPintegral}
\eea
where $\epsilon$ is the imaginary part of the Feynman propagator for the fermion running in the loop, rescaled by a mass dependent factor.  
In the case of an unstable fermion of total width $\Gamma_F$,  $\epsilon$ is a physical parameter: $\epsilon = m_F \Gamma_F / M_{Z'}^2$.
For a stable fermion $\epsilon\rightarrow 0_+$ is the usual Feynman propagator prescription.
While the remaining $x$-integral can be done analytically the result is not illuminating.  Instead, we compute the integral 
numerically as a function of $m_F$ and plot it in Figure~\ref{fig:FPintegral} for a few choices  of $M_{Z'}$.
We also compute the integral analytically in two limits
\be
I (r_h, r_F) =
 \begin{cases}
 \begin{aligned}
       & \;\; \frac{2 \, r_F}{1- r_h^2}\left[ 1+\left(\frac{2}{1- r_h^2}+\log\frac{ r_F^2}{ r_h}\right)\log r_h \right]~~~ &\mathrm{if}~~~   r_F\ll r_h &  ~~~. \\  
            & - \frac{1}{ 3 \, r_F }   
     ~~~  & \mathrm{if} ~~~  r_F\gg 1 & ~~~,     
 \end{aligned}
 \end{cases}
 \ee
 where we took into account that $r_h < 1$.

%
\vfil
\end{document}